\documentclass{llncs}

\usepackage{amsmath}
\usepackage{enumitem}
\usepackage[normalem]{ulem}
\usepackage{xspace}
\usepackage{hyperref}
\usepackage{cleveref}
\usepackage{url}

\usepackage{breakurl}
\usepackage{xcolor}
\usepackage{colortbl}
\usepackage{booktabs, makecell, tabularx,}
\usepackage{subcaption}
\usepackage{amssymb}
\usepackage[export]{adjustbox}

\newcommand{\pk}{\mathsf{pk}}
\newcommand{\sk}{\mathsf{sk}}

\newcommand{\addr}{\mathsf{addr}}
\newcommand{\tx}{\mathsf{tx}}

\newcommand{\channelid}{\mathsf{cid}}
\newcommand{\capacity}{\mathsf{C}}
\newcommand{\balance}{\mathsf{B}}
\newcommand{\inward}{\mathsf{in}}
\newcommand{\outward}{\mathsf{out}}
\newcommand{\fee}{\mathsf{fee}}
\newcommand{\amt}{\mathsf{amt}}

\newcommand{\PtoWSH}{\textsf{P2WSH}\xspace}

\newcommand{\onion}{\mathsf{onion}}

\newcommand{\fund}{\mathsf{fund}}
\newcommand{\local}{\mathsf{local}}
\newcommand{\remote}{\mathsf{remote}}
\newcommand{\htlc}{\mathsf{htlc}}

\newcommand{\testnet}{\textsf{testnet}\xspace}

\newcommand{\ipvf}{\textsf{IPv4}\xspace}
\newcommand{\ipvs}{\textsf{IPv6}\xspace}
\newcommand{\onionaddr}{\textsf{.onion}\xspace}

\newcommand{\throughputpernode}{\mathsf{T}_{x}}
\newcommand{\tUB}{\mathsf{throughput}_{\mathsf{big}}}
\newcommand{\tLB}{\mathsf{throughput}_{\mathsf{small}}}
\newcommand{\pUB}{\mathsf{lengths}_{\mathsf{long}}}
\newcommand{\pLB}{\mathsf{lengths}_{\mathsf{short}}}

\newcommand{\totalpay}{\mathsf{t_{pay}}}

\newcommand{\phead}[1]{\noindent{\bf\normalsize #1.}}

\newcommand\successprob[1]{\textrm{Pr}_{#1}^{\mathsf{succ}}}
\newcommand\failprob[1]{\textrm{Pr}_{#1}^{\mathsf{fail}}}
\newcommand\prob[1]{\textrm{Pr}_{#1}}
\newcommand\psucc{\mathsf{succ}}
\newcommand\pfail{\mathsf{fail}}

\newcommand{\btcMainnetSize}{300~GB}

\newcommand{\cdeckerStatsTotalChannelsMainnet}{174,378\xspace}
\newcommand{\cdeckerStatslosedChannelsMainnet}{135,850\xspace}
\newcommand{\cdeckerStatsTotalCapacityMainnet}{3315.18~BTC\xspace}

\usepackage[multiple]{footmisc}

\newif\ifsubmission
\usepackage[affil-it]{authblk}
\usepackage{cite}

\newcommand*\samethanks[1][\value{footnote}]{\footnotemark[#1]}

\begin{document}
\date{}

\title{An Empirical Analysis of Privacy in the Lightning Network}

\author{George Kappos\inst{1}\thanks{Both authors contributed equally.}, Haaroon Yousaf\inst{1}\samethanks, Ania Piotrowska\inst{1, 2}, Sanket Kanjalkar\inst{3}, Sergi Delgado-Segura\inst{4}, Andrew Miller\inst{3,5}, Sarah Meiklejohn\inst{1}}

\institute{University College London \and Nym Technologies \and University of Illinois Urbana-Champaign \and PISA Research \and IC3}

\maketitle

\begin{abstract}

Payment channel networks, and the Lightning Network in particular, seem to
offer a solution to the lack of scalability and privacy offered by Bitcoin and
other blockchain-based cryptocurrencies.
Previous research has focused on the scalability, availability, and
crypto-economics of the Lightning Network, but relatively little attention
has been paid to exploring the level of privacy it achieves in practice.
This paper presents a thorough analysis of the privacy offered by the Lightning
Network, by presenting several attacks that exploit publicly available 
information about the network in order to learn
information that is designed to be kept secret, such as how many coins a node
has available or who the sender and recipient are in a payment routed
through the network.  

\end{abstract}

\section{Introduction}\label{sec:introduction}

Since its introduction in 2008, Bitcoin~\cite{nakamoto2019bitcoin} has become
the most widely adopted cryptocurrency.
The decentralized and permissionless
nature of Bitcoin allows all users to join the network and avoids the need
for intermediaries and authorities who control the flow of money between them.
Instead, the validity of each transaction is verified by a consensus decision
made by the network participants themselves; valid transactions are then
recorded in the public blockchain.  The blockchain thus acts as a ledger
of all transactions that have ever taken place.

The need to broadcast transactions to all peers in the network and store them
in a permanent ledger, however, presents two problems for the longevity of
blockchain-based cryptocurrencies.  First, it imposes severe scalability
limitations: the Bitcoin blockchain today is over \btcMainnetSize, and Bitcoin can
achieve a throughput of only ten transactions per second.  Other
cryptocurrencies achieve somewhat higher throughputs, but there is an inherent
tradeoff in these broadcast-based systems between throughput and
security~\cite{gervais2016security,croman2016scaling}.  Second, the transparent
nature of the ledger means anyone can
observe the flow of coins, identify the counterparties to a transaction, and
link different transactions.  This has been shown most decisively for
Bitcoin~\cite{reid2013analysis, ron2013quantitative, androulaki2013evaluating,
meiklejohn2013fistful, spagnuolo2014bitiodine}, but this type of analysis
extends even to cryptocurrencies that were explicitly designed with privacy in
mind~\cite{kumar2017traceability,HH19,yu-monero,moser2018empirical,kappos2018empirical,
zcash-anon, biryukov2019privacy}.

The most promising solutions that have been deployed today to address the
issue of scalability are so-called ``layer-two''
protocols~\cite{gudgeon2019sok}, with the Lightning Network
(LN)~\cite{poon2016bitcoin} emerging as the most popular one since its launch
in March 2018.  In Lightning, pairs of participants use the Bitcoin blockchain
to open and close \emph{payment channels} between themselves.  Within a
channel, these two users can make arbitrarily many \emph{off-chain} payments
between themselves, without having to use the blockchain.  Beyond
a single channel, Lightning supports multi-hop payment routing, meaning even
participants who are not connected directly can still route payments through a
broader \emph{payment channel network} (PCN).  Nodes in the network are
incentivized to route payments by a fee they can charge for payments
they forward.

In addition to the promise it shows in improving scalability, Lightning
also seems to address the issue of privacy.  As we elaborate on in
Section~\ref{sec:background}, the nodes in the network and most of the channels
in the network are publicly known in order to build up the PCN (although some
channels may be kept private), as is the \emph{capacity} of a given channel,
meaning the maximum payment value that can be routed through it.  The
individual \emph{balances} associated with the channel, however,
are kept secret.
Furthermore, payments are not broadcast to all peers and are not stored in a
public ledger.
Even if a payment passes through
multiple channels, onion routing is used to ensure that each node on the path
can identify only its immediate predecessor and successor.

As is the case with ledger-based cryptocurrencies, however, the gap in
Lightning between the potential for privacy and the reality is significant,
as we show in this work.
In particular, we consider four main privacy properties promised by
LN~\cite{lightningnetworkbolts,malavolta2017concurrency}:

\begin{description}[leftmargin=0.2cm, topsep=0pt,itemsep=0pt]

\item[Private channels] should allow two nodes to share a channel but 
keep its existence, along with all of its information (capacity, participants, 
etc.), hidden from the rest of the network.  We explore
this property in Section~\ref{sec:private-channels} by presenting a heuristic 
that identifies on-chain funding of private channels 
and one or even both of the participants.

\item[Third-party balance secrecy] says that although the capacity of the
channel is public, the respective balances of the participants should remain
secret.  We explore this property in Section~\ref{sec:balance_attack} by
presenting and evaluating a generic method by which an active attacker (i.e.,
one opening channels with nodes in the network) can discover channel balances.

\item[On-path relationship anonymity] says that intermediate nodes routing
the payment should not learn which other nodes, besides their immediate
predecessor or successor, are part of the payment's route.
We explore this property in
Section~\ref{sec:path_attack}, where we leverage an LN simulator we developed
(described in Section~\ref{sec:simulator}) to evaluate the ability
of an intermediate node to infer the sender and recipient in payments that
it routes.

\item[Off-path payment privacy] says that any node not involved in routing a
payment should not infer any information regarding the routing nodes or the
payment value.  We explore this property in Section~\ref{sec:payment_attack}
by presenting and evaluating a method by which an active attacker can use the
ability to discover balances to form \emph{network snapshots}.  By comparing
consecutive network snapshots, the attacker can infer payments by
identifying where and by how much the balances of channels changed.

\end{description}

\subsection{Ethical considerations}

The attacks presented in Sections~\ref{sec:path_attack}
and~\ref{sec:payment_attack} are evaluated on a simulated network rather than
the live one,
but our attack in Section~\ref{sec:balance_attack} is evaluated on the
live test network.  As in related active attacks on
Bitcoin~\cite{koshy2014bitcoin,biryukov2014bitcoin,bogatyy2019grin}, we made
every effort to ensure that our attacks did not interfere with the normal
functioning of the network: the messages sent during the attack have no
abnormal effect and do not cost any money to process, and their volume is
relatively modest (we sent at most
24~messages per node we attacked).  We thus believe that they did not have
any long- or short-term destructive effect on the nodes that processed them.
We disclosed the results of this paper to the developers of the three main LN
clients and the liquidity provider Bitrefill in February 2020, and have 
discussed the paper with the Lightning developers since then.

\subsection{Related work}\label{sec:related}

We consider as related all research that focuses on the Lightning Network,
particularly as it relates to privacy.
Most of the previous research has focused on the scalability, utility and crypto-economic
aspects of LN~\cite{khan2019lightning, branzei2017charge, khalil2017revive,
werman2018avoiding, malavolta2017concurrency}, or on
its graph properties~\cite{seres2019topological, martinazzi2019evolution}.
Rohrer et al.~\cite{rohrer2019discharged} study the susceptibility of LN
to topology-based attacks, and
Tochner et al.~\cite{tochner2019hijacking} present a DoS attack that
exploits how multi-hop payments are routed.
Among other findings, they show that the ten most central nodes can disrupt
roughly 80\% of all paths using their attack.  
P\'{e}rez-Sol\`{a} et al.~\cite{lockdown} present an attack that diminishes the capacity
of a node's channels, preventing it from participating in the network.
Tikhomirov et al.~\cite{cryptoeprint:2020:303} show how a wormhole attack prevents
honest intermediaries from participating in routing payments.

In terms of privacy, Malavolta et al.~\cite{malavolta2019multihop} identify a
new attack exploiting the locking mechanism, which allows dishonest users to
steal payment fees from honest intermediaries along the path.  They propose
anonymous multi-hop locks as a more secure option.  
Nowatowski and T{\o}n~\cite{nowostawski2019evaluating} study various
heuristics in order to identify Lightning transactions
on the Bitcoin blockchain.  Concurrently to our work, Romiti et 
al.~\cite{romiti2020crosslayer} developed several
heuristics to link Bitcoin wallets to Lightning entities.  One of their 
heuristics is similar to the \emph{tracing heuristic} we develop in 
Section~\ref{sec:private-channels}, but their goal is to create augmented 
Bitcoin clustering methods rather than identify private channels.
As we describe further in Section~\ref{sec:balance_attack}, others have performed balance discovery attacks~\cite{herrera2019difficulty,nisslmueller2020active,tikhomirov2020probing}.
The main limitation of these attacks is that they rely on specifics of the 
error messages the attacker receives, so may easily become irrelevant as the
network evolves.  We overcome this
limitation by presenting a \emph{generic} attack (in
Section~\ref{sec:balance_attack}), as well as investigating the
implications of such an attack more broadly (in
Section~\ref{sec:payment_attack}).  

B{\'e}res et al.~\cite{beres2019cryptoeconomic} look briefly
at the question of finding the sender and recipient of a payment.
Similarly to our work, they develop an LN
traffic simulator based on publicly available network snapshots and
information published by certain node owners. 
Their work considers only single-hop payments, however, and does not look at
other privacy properties.  
There are a number of other Lightning network studies that use a network
simulator~\cite{branzei2017charge,
conoscenti2018cloth, engelmann2017towards, zhang2019cheapay}.
Several of these simulators were used to perform economic analysis of the Lightning
network~\cite{branzei2017charge,engelmann2017towards,zhang2019cheapay},
while the CLoTH simulator~\cite{conoscenti2018cloth} provides only performance statistics
(e.g., time to complete a payment, probability of payment failure, etc.).
However, all of those simulators make several simplifying assumptions about the topology, path selection
algorithm, and distribution of payments.  As such, they are not suitable for
an analysis of its privacy properties.

\section{Background}\label{sec:background}

In order to open a Lightning \emph{channel}, two parties deposit 
bitcoins into a 2-of-2 \emph{multi-signature} address, meaning any
transaction spending these coins would need to be signed by both of them.  These
funds represent the channel \emph{capacity}; i.e., the maximum amount
of coins that can be transferred via this channel.
Once a channel is established, its participants  
can use it to exchange arbitrarily
many payments, as long as either has a positive balance.  They can also 
close the channel using a Bitcoin transaction that sends them their respective 
balances from the 2-of-2 multi-signature address.

Most users, however, are not connected directly, so instead need to 
\emph{route} their payments through
the global Lightning Network.  %
Here, nodes are identified by public keys, and edges represent
channels, which are publicly associated with a 
\emph{channel identifier} $\channelid$, the channel capacity $C$,
and a \emph{fee} $\fee$ that is charged for routing payments via this
channel.
Privately, edges are also implicitly associated with the \emph{inward} and
\emph{outward}
\emph{balances} of the channel.  %
Except for \emph{private} channels, which are revealed only at the time of
routing, the topology of this network and its public labels are known to every
peer.  
\newcommand\lndetails{
\subsection{Channel management}\label{sec:managment}

In this section, we describe in detail the channel management operations,
including its opening and closing, and channel state updates.

\subsubsection{Opening a channel}\label{sec:opening}

For Alice and Bob to create a channel between them,
they must fund the channel by forming a \emph{funding transaction} $\tx_\fund$.
Let's assume that Alice is the one funding the channel.\footnote{In general,
either one of the parties funds the channel, or both of them.}
The funding transaction consists of one input, which is a UTXO associated with one of Alice's
addresses, and one output, which is a $2$-of-$2$ multisig address representing both
Alice and Bob's public keys. This means that both Alice and Bob must provide
their signatures in order to release the funds.
The amount sent to the multisig address is the initial capacity $\capacity$ of
the channel.

It is important that Alice does not just put this transaction on the Bitcoin
blockchain; otherwise, her coins may be locked up if Bob refuses to provide a
signature.  Instead, she and Bob first go on to form two \emph{commitment
transactions}, one for each of them, which represent their agreement on the
current state of the channel.  Each commitment transaction has the 2-of-2
multisig as input, and has two outputs: $\local$ and $\remote$.
The transaction $\tx_{A,i}$, representing Alice's view of state $i$,
essentially sends her current balance to $\local$ and sends Bob's current
balance to $\remote$.  Likewise, $\tx_{B,i}$ sends Bob's balance to
$\local$ and Alice's balance to $\remote$.

The $\remote$ output is simply the address of the other involved party
(so $\addr_B$ in $\tx_{A,i}$ and $\addr_A$ in
$\tx_{B,i}$), but the $\local$ output is more complicated.  Instead, the
$\local$ output in $\tx_{B,i}$ is
encoded with some timeout $t$, and awaits some potential input $\sigma$.  If
the timeout $t$ has passed, then the funds go to $\addr_B$ as planned.
Otherwise, if a valid \emph{revocation signature} $\sigma$ is provided before
time $t$, the funds go to $\addr_A$.  This means that the revocation signature
allows one party to claim the entire capacity of the channel for themselves.
As we explore fully in Section~\ref{sec:updating-state}, this is used to
disincentive bad behavior in the form of publishing old states.

To create a channel, Alice thus creates the transaction $\tx_{B,0}$, sending
$\capacity$ to $\remote$ and $0$ to $\local$ (since so far she has supplied
all the funds for the channel).  She signs this transaction and sends her
signature to Bob, who signs it as well.  Bob then forms
$\tx_{A,0}$, sending $0$ to $\remote$ and $\capacity$ to $\local$, and sends
his signature on this to Alice.  Alice then signs it and publishes
$\tx_\fund$ to the Bitcoin blockchain.

At this point, Alice and Bob both have valid transactions, meaning
transactions that are signed by both parties in the input multisig, which is
itself the output of a transaction on the blockchain (i.e., the funding
transaction).  Either of them could thus publish their transaction to the
blockchain to close the channel.  Alternatively, if they mutually agree to
close the channel and want to avoid having one of them wait until the
timeout to claim their funds, they could update to a new state without the
timeout in $\local$ and publish that.

\subsubsection{Updating a channel state}\label{sec:updating-state}

Once both Alice and Bob have signed $\tx_{A,0}$ and $\tx_{B,0}$, they have
agreed on the \emph{state} of the channel, which
represents their respective balances $\balance_A$ and $\balance_B$.
In particular, the amount sent to $\local$ in $\tx_{A,0}$ and to $\remote$ in
$\tx_{B,0}$ represents Alice's balance, and similarly the amount sent to
$\local$ in $\tx_{B,0}$ and to $\remote$ in $\tx_{A,0}$ represents Bob's
balance.  The question is now how these transactions are updated when one
of them wants to pay the other one, and thus these balances change.

In theory, this should be simple: Alice and Bob can repeat the same process
as for the creation of these initial commitment transactions, but with the
new balances produced by the payment.  The complicating factor, however, is
if the old commitment transactions are still valid after the creation of the
new ones, then one of them might try to revert to an earlier state by
broadcasting an old commitment transaction to the Bitcoin blockchain.
For example, if Alice pays Bob in exchange for some service, then it is important
that once the service has been provided, Alice cannot publish an old commitment
transaction claiming her (higher) balance before she made the payment.

This is exactly the role of the \emph{revocation signature} introduced earlier.
In addition to exchanging signatures on the new transactions $\tx_{A,i+1}$ and
$\tx_{B,i+1}$, Alice and Bob also exchange the revocation secret keys $\sk_{A,i}$
and $\sk_{B,i}$ that correspond to the public keys $\pk_{A,i}$ and $\pk_{B,i}$
used in $\tx_{A,i}$ and $\tx_{B,i}$.
These public keys are derived from base public keys $\pk_A$ and $\pk_B$ such that
(1) both Alice and Bob can compute $\pk_{A,i}$ and $\pk_{B,i}$ for any state
$i$, thus can independently form $\tx_{A,i}$ and $\tx_{B,i}$ as long as they
know the right amounts, and (2) the corresponding secret keys $\sk_{A,i+1}$
and $\sk_{B,i+1}$ are completely unknown until one party reveals them to the
other.  Thus, up until they exchange revocation keys, they can broadcast the
transactions for state $i$ to the Bitcoin blockchain as a way to close the
channel.  Once they exchange the secret keys, however, Bob can form a
valid revocation signature and claim all of Alice's funds in $\local$ if
she broadcasts $\tx_{A,i}$ (it is indeed only Alice who can broadcast
a valid $\tx_{A,i}$ as only she has both her and Bob's signatures
on it). At this point the new channel state $i+1$ is confirmed and
Bob can safely perform his service for Alice.

\subsubsection{Closing a channel}\label{sec:closing}

As long as either Alice or Bob has a positive balance, they can send payments
to the other, knowing that if there is a disagreement they can settle their
previously agreed channel state onto the blockchain, as described in the
previous section.  If there is no dispute, and both Alice and Bob agree to
close the channel, they perform a \emph{mutual close} of the channel.  This
means providing a signature that authorizes a \emph{settlement transaction}.
}
When routing a payment, the sender (Alice) uses \emph{onion routing} to hide 
her relationship with the recipient (Bob).
Alice selects the entire path to Bob (\emph{source routing}),
based on the capacities and fees of the channels between them.  
The eventual goal is that each
intermediate node on this path forwards the payment to its successor, 
expecting that its
predecessor will do the same so its balance will not change.  The nodes
cannot send the money right away, however, because it may be the case that the
payment fails.
To thus create an intermediate state, 
LN uses \emph{hashed time-lock contracts}
(\emph{HTLCs}), which allow for time-bound conditional payments.
In summary, the protocol follow five basic steps to have Alice pay Bob:

\begin{description}[itemsep=0pt,leftmargin=0.2cm, topsep=0pt]

\item[1. Invoicing] Bob generates a secret $x$ and computes the hash $h$ of
it.  He issues an \emph{invoice} containing $h$
and some payment amount $\amt$, and sends it to Alice.

\item[2. Onion routing]%
Alice picks a path
$A\rightarrow U_1\rightarrow\cdots\rightarrow
U_n\rightarrow B$.  %
Alice then forms a
Sphinx~\cite{danezis2009sphinx} packet destined for Bob and routed
via the $U_i$ nodes.
Alice then sends the outermost onion packet $\onion_1$ to $U_1$.

\item[3. Channel preparation] Upon receiving $\onion_i$ from $U_{i-1}$, $U_i$
decodes it to reveal:
$\channelid$, which identifies the next node $U_{i+1}$,
the amount $\amt_i$ to send them, a timeout $t_i$, and the packet
$ \onion_{i+1}$ to forward to $U_{i+1}$.
Before sending $\onion_{i+1}$ to $U_{i+1}$, $U_i$
and $U_{i-1}$ \emph{prepare} their channel by updating their
intermediate state using an HTLC, which 
ensures that if $U_{i-1}$ does not provide $U_i$ with the pre-image of $h$
before the timeout $t_i$, $U_i$ can claim a refund of their payment.
After this is done, $U_i$ can send $\onion_{i+1}$ to
$U_{i+1}$.

\item[4. Invoice settlement] Eventually, Bob receives $\onion_{n+1}$ from $U_n$
and decodes it to find $(\amt,t,h)$.  If $\amt$ and $h$ match what
he put in his invoice, he sends the invoice pre-image $x$ to $U_{n-1}$ in order
to redeem his payment of $\amt$. This value is in turn sent backwards along
the path.

\item[5. Channel settlement] At every step on the path, $U_i$ and $U_{i+1}$
use $x$ to \emph{settle} their channel; i.e., to confirm the updated
state reflecting the fact that $\amt_i$ was sent from $U_i$ to
$U_{i+1}$ and thus that $\amt$ was sent from Alice to Bob.

\end{description}

\section{Blockchain Analysis}\label{sec:on-chain}

\subsection{Data and measurements}\label{sec:measurements}

The Lightning network can be captured over time by periodic snapshots of the
public network graph, which provide ground-truth data about nodes
(identifiers, network addresses, status, etc.) and their channels
(identifiers, capacity, endpoints, etc.).  To obtain a comprehensive set of
snapshots, we used data provided to us by (1) our own \textsf{lnd} client,
(2) one of the main c-lightning developers, and (3) scraped user-submitted
(and validated) data from 1ML\footnote{\url{https://1ml.com/}} and LN
Bigsun.\footnote{\url{https://ln.bigsun.xyz/}}
To analyze on-chain transactions, we also ran a full Bitcoin node, using
the BlockSci tool~\cite{Kalodner2017BlockSciDA} to parse and analyze the raw
blockchain data.

Our LN dataset included the hash of the Bitcoin transaction
used to open each channel.  By combining this with our blockchain data, we
were thus able to identify when channels closed and how their funds were
distributed.
In total, we identified \cdeckerStatsTotalChannelsMainnet channels, of
which \cdeckerStatslosedChannelsMainnet had closed with a total capacity of \cdeckerStatsTotalCapacityMainnet.
Of the channels that closed,
69.22\% were claimed by
a single output address (i.e., the channel was completely unbalanced at the
time of closure), 29.01\% by two output addresses, and
1.76\% by more than two outputs.

\subsection{Private channels}\label{sec:private-channels}

Private channels provide a way for two Lightning nodes to create a channel but
not announce it to the rest of the network.  In this section, we seek to
understand the extent to which private channels can nevertheless be
identified, to understand their privacy limitations as well as the scope of
our remaining attacks on the public network. 

We first provide an upper bound on the number of private channels using a
\textit{property heuristic}, which identifies Bitcoin transactions that seem
to represent the opening and closing of channels but for which we have no
public channel identifier.  

\subsubsection{Property Heuristic}\label{sec:property_heuristic_appendix}
To align with our LN dataset, we first looked for all
Bitcoin transactions that (1) occurred after January 12, 2018 %
and (2) before September 7, 2020, %
and (3) where one of the outputs was a \PtoWSH address (which LN channels have
to be, according to the specification).  We identified
3,500,312 transactions meeting these criteria, as compared with the 174,378
public channels opened during this period.
We then identified several common
features of the known opening transactions identified from our dataset:
(i) 99.91\% had at most two outputs, which
likely represents the funder creating the channel and
sending themselves change;
(ii) 99.91\% had a single \PtoWSH output address;
(iii) 99.85\% had a \PtoWSH output address that received at most 16,777,215
satoshis, which at the time of our analysis is the maximum capacity of an
LN channel;
(iv) 99.99\% had a \PtoWSH output that appeared at most once as both an input and
output, which reflects its ``one-time'' usage as a payment channel and not as
a reusable script; and
(v) 99.99\% were funded with
either a \textsf{WitnessPubHeyHash} address or \textsf{ScriptHash} address.

By requiring our collected transactions to also have these features and excluding
any transactions involved in opening or closing public channels, we
were left with 267,674 potential transactions representing the opening of
private channels.  %
If the outputs in these transactions had
spent their contents (i.e., the channel had been closed), then we were further
able to see how they did so, which
would provide better evidence of whether or not they were associated with
the Lightning Network.
Again, we identified the following features based on known closing
transactions we had from our network data:
(i) 100\% had a non-zero sequence number, as required by the Lightning
specification~\cite{lightningnetworkbolts};
(ii) 100\% had a single input that was a 2-of-2 multisig address, again as required
by the Lightning design; and %
(iii) 98.24\% had at most two outputs, which reflects the two participants
in the channel.

By requiring our collected opening transactions to also have a closing
transaction with these three features, we were left with 77,245 pairs of
transactions that were potentially involved in opening and closing private
channels. %
Again, this is just an upper
bound, since there are other reasons to use 2-of-2 multisigs in this way that have
nothing to do with Lightning.

We identified 77,245 pairs of transactions that were potentially
involved in opening and closing private channels, but likely has a high false 
positive rate. We thus developed a \textit{tracing heuristic}, which
follows the ``peeling chain''~\cite{meiklejohn2013fistful} initiated at the
opening and closing of public channels to identify any associated private
channels.

\subsubsection{Tracing heuristic.}\label{tracingheuristic}

We next look not just at
the properties of individual transactions, but also at the flow of bitcoins
between transactions.  In particular, we observed that it was common for users
opening channels to do so in a ``peeling chain''
pattern~\cite{meiklejohn2013fistful}.  This
meant they would (1) use the change in a channel opening
transaction to continue to create channels
and (2) use the outputs in a channel closing transaction to open new channels.
Furthermore, they would often (3) co-spend change with closing outputs;
i.e., create a channel opening transaction in which the input addresses were
the change from a previous opening transaction and the output from a previous
closing one.

By systematically identifying these operations, we were able to link together
channels that were opened or closed by the same Lightning node 
by following the peeling chain both forwards and backwards.  Going
backwards, we followed each input until we hit a transaction that did not seem
to represent a channel opening or closure, according to our property
heuristic.  Going forwards, we identified the change address in a transaction,
again using the property heuristic to identify the channel creation address
and thus isolate the change address as the other output, and continued
until we hit one of the following: (1) a transaction with no change output or
one that was unspent, meaning we could not move forwards, or (2) a transaction
that did not satisfy the property heuristic.  We also did this for all of the
outputs in a known channel closing transaction, to reflect the second pattern
identified above.

We started with the 174,378 public channels identified in our LN dataset.  By
applying our tracing heuristic, we ended up with 27,386 additional
channel opening transactions.
Of these, there were 27,183 that fell within the same range of blocks as the
transactions identified by our property heuristic.

Using the tracing heuristic, however, not only identified private channels but
also allowed us to
cluster together different channels (both public and
private), according to the shared ownership of transactions
within a peeling chain~\cite{meiklejohn2013fistful}.
To this end, we first clustered together different channels according to their
presence in the same peeling chain, and then looked at the public channels
within each cluster and
calculated the common participant, if any, across their endpoints.
If there was a single common participant, then we could confidently tag them
as the node responsible for opening
all of these channels.

In order to find the other endpoint of each private channel, we followed the
closing outputs of the channel's closing transaction, whenever applicable,
leveraging the second and third observed patterns in the tracing heuristic.
In particular, when a closing output was spent in
order to open a new channel, we performed the same
clustering operation as earlier.
We failed to identify the second participant in each channel only when
the channel was still open, the channel was closed but the closing output was
still unspent, or the closing output was used for something other than Lightning.

Out of the 27,183 transactions we identified as representing the opening of
private channels, we were able to identify both participants in 2,035
(7.5\%), one participant in 21,557 (79.3\%), and no participants in 3,591
(13.2\%).  Our identification method applies equally well, however, to public
channels.
We were able to identify
the opening participant for 155,202 (89.0\%) public channels.  Similarly, for
the public channels that were already closed, which represent 185,860 closing
outputs, we were able to associate 143,577 (77.25\%) closing outputs with a
specific participant.

\section{Balance Discovery}\label{sec:balance_attack}

Previous attacks designed to discover the balances associated with individual
channels (as opposed to just their capacity)~\cite{herrera2019difficulty,nisslmueller2020active,tikhomirov2020probing} 
exploited debug information as an \emph{oracle}.  In these attacks, an attacker 
opens a channel with a node and 
routes a fake payment hash, with some associated amount $\amt$, through its 
other channels.  Based on the error messages received and performing a binary
search on $\amt$ (i.e., increasing $\amt$ if the payment went through and
decreasing it if it failed), the attacker efficiently determines the exact 
balance of one side of the channel.  
In this section we perform a new \emph{generic} attack on the LN \testnet.  As
compared to previous attacks, our attacker must run two nodes rather than one.
If error messages are removed or made generic in the future, however, our 
attack would continue to work whereas previous attacks would not.

\subsubsection{The attack.}
In our attack, an attacker running nodes $A$ and $D$ needs to form a path
$A\rightarrow B\rightarrow C\rightarrow D$, with the goal of finding the
balance of the channel $B\rightarrow C$.  This means our attacker needs to run
two nodes, one with a channel with outgoing balance ($A$), and one with a 
channel with incoming balance ($D$).  Creating the channel $A\rightarrow B$ is
easy, as the attacker can just open a channel with $B$ and fund it themselves.
Opening the channel $C\rightarrow D$ is harder though, given that the attacker
must create incoming balance.

Today, there are two main options for doing this.  First, the
attacker can open the channel $C\rightarrow D$ and fund it themselves, but
assign the balance to $C$ rather than to $D$ (this is called
funding the ``remote balance'').  This presents the risk, however, that $C$ 
will immediately close the channel and take all of its funds. We call this 
approach unassisted channel opening. 
The second option is to use a
\emph{liquidity provider} (e.g.,
Bitrefill\footnote{\url{https://www.bitrefill.com/}} or
LNBIG\footnote{\url{http://lnbig.com/}}), which is a service that
sells channels with incoming balance.\footnote{Bitrefill, for example, sells a channel
	with an incoming balance of 5000000~satoshis (the equivalent at the time of
	writing of 493.50~USD) for 8.48~USD.}
We call this assisted channel opening. 

Once the attacker has created the channels $A\rightarrow B$ and $C\rightarrow
D$, they route a random payment hash $H$ to $D$, via $B$ and $C$, with some 
associated amount $\amt$.  
If $D$ receives $H$, this means the 
channel from $B$ to $C$ had sufficient balance to route a payment of amount 
$\amt$.  If $D$ did not receive $H$ after some timeout, the attacker can assume 
the payment failed, meaning $\amt$ exceeded the balance from $B$ to $C$.  Either 
way, the attacker can (as in previous attacks) repeat the
process using a binary search on $\amt$.
Eventually, the attacker discovers the balance of the channel as the
maximum value for which $D$ successfully receives $H$.

To a certain extent, this attack generalizes even to the case in which there is
more than one intermediate channel between the two attacker nodes.  In this more
general case, however, the above method identifies the bottleneck balance
in the entire path, rather than the balance of 
an individual channel.
In the event of a payment failure though, the current C-lightning and LND 
clients return an \emph{error index}, which is the position of the
node at which the payment failed.  This means that an attacker would know
exactly where a payment failed along a longer path.  We chose not to use this
index when implementing our attack, in order to keep it fully generic and to
test just the basic version, but leave an attack that does use this index as 
interesting future research.

\subsubsection{Attack results.}

We performed this attack on \testnet on September 3 2020.  We ran two LN nodes
and funded all our
channels (unassisted), both locally and remotely, which required a slight
modification of the client (as fully funding a remote channel is restricted by
default).  We opened channels with every accessible 
node in the network. At the time of the attack there were 3,159 nodes and 
9,136 channels, 
of which we were able to connect to 103 nodes and 
attack 1,017 channels.  %
We were not able to connect to a majority of the overall nodes, which 
happened for a variety of reasons: some nodes did not publish an IPv4 address, 
some were not online, some had their advertised LN ports closed, and some 
refused to open a channel.

Of these 1,017 channels, we determined the balance of 568. %
Many (65\%) of the channels were fairly one-sided, meaning the balance of the 
attacked party was 70\% or more of the total capacity.
We received a variety of errors for the channels where we were unsuccessful,
such as \textit{TemporaryChannelFailure}, 
or we timed out as the client took 
more than 30 seconds to return a response.

We did not carry out the attack on mainnet due to cost and ethical
considerations, but believe it likely that the attack would perform better
there. %
This is because there is no cost for forgetting to close open channels on testnet or maintain a node, whereas
 on mainnet a user is incentivized by an opportunity cost (from fees) to ensure a node is maintained and its channels are active.

\subsubsection{Attacker cost.}
In our experiment we used testnet coins, which are of essentially no value, so 
the monetary cost for us to perform this attack was negligible. To understand
the practical limitations of this attack, however, we estimate the minimum 
cost on mainnet. 
When creating the outgoing channel $A\rightarrow B$, the attacker must pay for
the opening and closing transaction fees on the Bitcoin blockchain.  At the
time of our attack, this was 0.00043~BTC %
per transaction.  They must 
also remotely fund the recipient node with enough \emph{reserve satoshis} to 
allow the forwarding of high payments,  
which at present are 1\% of the channel capacity.
To create the incoming channel $C\rightarrow D$, the attacker 
can use liquidity providers like Bitrefill, who at the time of writing 
allow users to buy channels with 0.16~BTC incoming capacity for 0.002604~BTC.  

Purchasing the cheapest incoming liquidity available today would cost the
attacker 0.00086~BTC and 0.005~BTC on hold, enabling routes to 4,811 channels
(with a total capacity of 45~BTC). This would require opening 2,191 channels 
with a maximum channel capacity of 0.04998769~BTC.  In total, this would
require the attacker to spend 1.097~BTC and put 109.53~BTC on hold.

\section{Path Discovery}\label{sec:path_attack}

We now describe how an \emph{honest-but-curious} intermediate node involved in
routing the payment can infer information regarding its path, and in
particular can identify the sender and recipient of the payment.
Our strategy is similar to a passive variant of a predecessor attack~\cite{DBLP:journals/tissec/WrightALS04}
proposed against the Crowds~\cite{DBLP:journals/tissec/ReiterR98} anonymous
communication network. Our strategy can be further extended by analyzing the
sparse network connectivity and limited number of potential paths due to channel
capacity.

In contrast to previous work~\cite{beres2019cryptoeconomic}, we consider not only
single-hop routes but also routes with multiple intermediate nodes. The only
assumption we make about the adversary's intermediate node is that it keeps
its channel balanced, which can be easily done in practice. %

We define $\prob{S}$ and $\prob{R}$ as the probability that the adversary
successfully discovers, respectively, the sender and recipient in a payment.
Following our notation, Béres et al.\ claim, based on their own simulated results,
that $\prob{S} = \prob{R}$ ranges from $0.17$ to $0.37$ depending on parameters
used in their simulation. We show that this probability is actually a lower bound,
as it does not take into account multiple possible path lengths or the chance that
a payment fails (their simulation assumes that all payments succeed on the first try).

The strategy of our honest-but-curious adversary is simple: they always guess
that their immediate predecessor is the sender.
In other words, if we define $H$ as the adversary's position along the path, they
always assume that $H=1$. Similarly, they always guess that their immediate
successor is the recipient.  We focus on the probability of
successfully guessing the sender; the probability of succesfully guessing
the recipient can be computed in an analogous way.

\vspace{2mm}
\phead{Successful payments}
We start by analyzing the success probability of this adversary in the case of
a successful payment, which we denote as $\successprob{S}$.
We define as $\Pr[L = \ell]$ the probability of a path being of length $\ell$,
and as $ \Pr[H = h~|~L = \ell]$ the probability that the adversary's node is at
position $h$ given that the path length is $\ell$. According to the Lightning
specification~\cite{lightningnetworkbolts}, the maximum path length is $20$.
By following the strategy defined above, we have that
\begin{align*}
\successprob{S} &= \sum_{n=3}^{20} \Pr[L=\ell~|~\psucc] \cdot
\Pr[H=1~|~L=\ell, \psucc] \\
&=  \Pr[L = 3~|~\psucc] \\
&~~~+ \sum_{n=4}^{20}  \Pr[L=\ell~|~\psucc] \cdot
\Pr[H=1~|~L=\ell,\psucc]
\end{align*}
since $\Pr[H=1~|~L= 3, \psucc]=1$ given that the adversary is the only
intermediate node in this case.  Hence, $ \Pr[L = 3~|~\psucc]$ is a
lower bound on $\prob{S}$.

To consider the overall probability, we focus on the conditional probabilities
$\Pr[H=1~|~L=\ell, \psucc]$. If all nodes form a
clique,\footnote{This would rather be a clique excluding a link between the
sender and
recipient, since otherwise they would presumably use their channel directly.}
 then it would be almost equally probable for any node to be in any hop
position $H=h$. (The only reason the distribution is not entirely uniform is
that some channels may be chosen more often than others, depending on the
relative fees they charge, but an adversary could choose fees to
match its neighbors as closely as possible.)  In this case then, the
probability that $H=1$ is just $1 / (\ell-2)$.

\vspace{2mm}
\phead{Failed payments}
Similarly, in case the payment fails, we define the probability
$\failprob{S}$ as
\begin{align*}
\failprob{S} = \sum_{\ell=3}^{20} \Pr[L=\ell~|~\pfail] \cdot
    \Pr[H=1~|~L=\ell, \pfail].
\end{align*}
This is the same formula as for $\successprob{S}$ so far, but we know that
$\Pr[L=3~|~\pfail] = 0$, since if the adversary is the only
intermediate node the payment cannot fail.
Furthermore, the conditional
probability $\Pr[H=1~|~L=\ell, \pfail]$ is different from the probability
$\Pr[H=1~|~L=\ell,\psucc]$, as the fact that a payment failed reveals
information to the adversary about their role as an intermediate node.
In particular, if an intermediate node successfully forwards the payment to
their successor but the payment eventually fails, the node learns that their
immediate successor was not the recipient and thus that the failed path was of
length $L\geq 4$ and their position is not $L-1$.
This means that $\Pr[L=\ell~|~\pfail]$ becomes
$\Pr[L=\ell~|~\pfail,\ell \geq 4]$.  We thus get
\begin{align*}
\failprob{S} &= \Pr[L=4~|~\pfail, \ell \geq 4]  \\
    &~~+ \sum_{\ell=5}^{20}  \Pr[L=\ell~|~\pfail] \cdot \Pr[H=1~|~L=\ell,
\pfail].
\end{align*}
This gives $\Pr[L=4~|~\pfail, \ell \geq 4]$ as a lower bound in the case of a
failed payment.  As we did in the case of successful payments, we assume a
clique topology as the best case for this adversary's strategy, in which their
chance of guessing their position is $1 / (\ell-3)$ (since they know they are
not the last position).  We thus obtain
\begin{align*}
\failprob{S} &= \Pr[L=4~|~\pfail, \ell \geq 4]  + \sum_{\ell=5}^{20}
    \Pr[L=\ell~|~\pfail] \cdot \frac{1}{\ell-3}.
\end{align*}

\subsection{Lightning Network simulator}\label{sec:simulator}

In order to investigate the success of this on-path adversary, we need
measurements that it would require
significant resources to obtain from the live network, such as the average
path length for a payment.  Given the financial and ethical concerns this
would raise, we make the same decision as in previous work~\cite{beres2019cryptoeconomic, branzei2017charge,
conoscenti2018cloth, engelmann2017towards, zhang2019cheapay} to develop a
Lightning network simulator to perform our analysis.
We implemented our
simulator in 2,624 lines of Python~3 and will release it as open-source software.

\label{sec:network_simulation}

\vspace{2mm}
\phead{Network topology}
As mentioned in~\Cref{sec:background}, we represent the network as a graph $G = (V, E)$.
We obtain the information regarding $V$ and $E$ from the snapshots we
collected, as described in Section~\ref{sec:measurements}, which also include
additional information such as capacities and fees.  The network
topology view of our simulator is thus an exact representation of the actual
public network.

\vspace{2mm}
\phead{Geolocation}
Nodes may publish an \ipvf, \ipvs
or \onionaddr address, or some combination of these.  If a node advertised an
\ipvf or \ipvs address then we used it to assign this node a corresponding
geolocation. This enabled us to accurately simulate TCP delays on the packets
routed between nodes, based on their distance and following previous
studies~\cite{gervais2016security} in using the global IP latency from
Verizon.\footnote{\url{https://enterprise.verizon.com/terms/latency/}}
For nodes that published only a \onionaddr, we assign
delays according to the statistics published by Tor metrics,
given the higher latency associated with the Tor network.\footnote{\url{https://metrics.torproject.org/onionperf-latencies.html}}

\vspace{2mm}
\phead{Path selection}
As discussed in~\Cref{sec:background}, the route to the destination in LN is
constructed solely by the payment sender.  All clients generally aim to find
the shortest path in the network, meaning the path with the lowest amount of
fees.  As shown by Tochner et al.~\cite{tochner2019hijacking}, however, both
the routing algorithm and the fee calculation differ across the three
main choices of client software: \textsf{lnd},
\textsf{c-lightning}, and \textsf{eclair}.  We could not easily extract or
isolate the routing algorithms from these different implementations, so chose
to implement all three
versions of the path finding algorithm ourselves.
We did this using Yen's $k$-shortest path algorithm~\cite{yen1970algorithm}
and the \textsf{networkx} Dijkstra's SPF algorithm.\footnote{\url{https://networkx.github.io/documentation/stable/reference/algorithms/shortest_paths.html}}

\vspace{2mm}
\phead{Software versions}
Our collected snapshots did not include information about software versions,
so we scraped the \textsf{Owner Info} field for each node listed on the 1ML
website. Although in 91\% of the cases this field is empty,
the results allow us to at least estimate the distribution of the client software.
We obtained information about $370$ nodes
and found that $292$ were \textsf{lnd}, $54$ were \textsf{c-lightning}, and
$24$ were \textsf{eclair}.  We randomly assign software versions to the
remaining nodes in the network according to this distribution, and then modify
the \textsf{weight} function in the path finding algorithm according to
the software version.

\subsubsection{Payment parameters}
\label{sec:simulating_payments}

Our first parameter, $\totalpay$, represents the total
daily number
of payments happening in LN. For this, we use an estimate from
LNBIG~\cite{lnbigarticle}, the largest node that holds more than 40\% of the
network's total capacity at the time of writing. According to LNBIG, the total
number of routed transactions going through the network is 1000-1500 per day,
but this does not take into account the payments performed via direct channels.
Given this estimation, we use two values for $\totalpay$: 1000, representing a
slight underestimate of today's volume, and 10,000, representing a potential
estimate for the future of LN.

We also define as $\mathsf{endpoints}$ the parameter that determines the
sender and the recipient of a payment. We define two values for this
parameter: \emph{uniform}, which means that the payment participants are chosen
uniformly at random, and \emph{weighted}, which means the participants are chosen
randomly according to a weighted distribution that takes into account their
number of direct channels (i.e., their degree).
Similarly, we use $\mathsf{values}$ to determine the values of payments.
When $\mathsf{values}$ is \emph{cheap}, the payment value is the smallest value
the sender can perform, given its current balances.  When $\mathsf{values}$ is
\emph{expensive}, the payment value is the biggest
value the sender can send.

\subsection{Simulation results}
\label{sec:simulation_results}

Given the parameters $\totalpay$, $\mathsf{endpoints}$, and $\mathsf{values}$,
we ran two simulation instances, with the goal of finding the worst-case and
best-case scenarios for the on-path adversary.  Based on the respective
probabilities for $\successprob{S}$ and $\failprob{S}$, we
can see that the worst case is when the path is long and the payment is
likely to succeed, while the best case is when the path is short and the
payment is likely to fail.  Since the
total volume $\totalpay$ does
not affect the path length, we use $\totalpay = 1000$ for both instances.
Each simulation instance was run using the network and node parameters scraped
on September 1, 2020.

\begin{figure}[t!]
  \centering
	\includegraphics[width=0.6\linewidth]{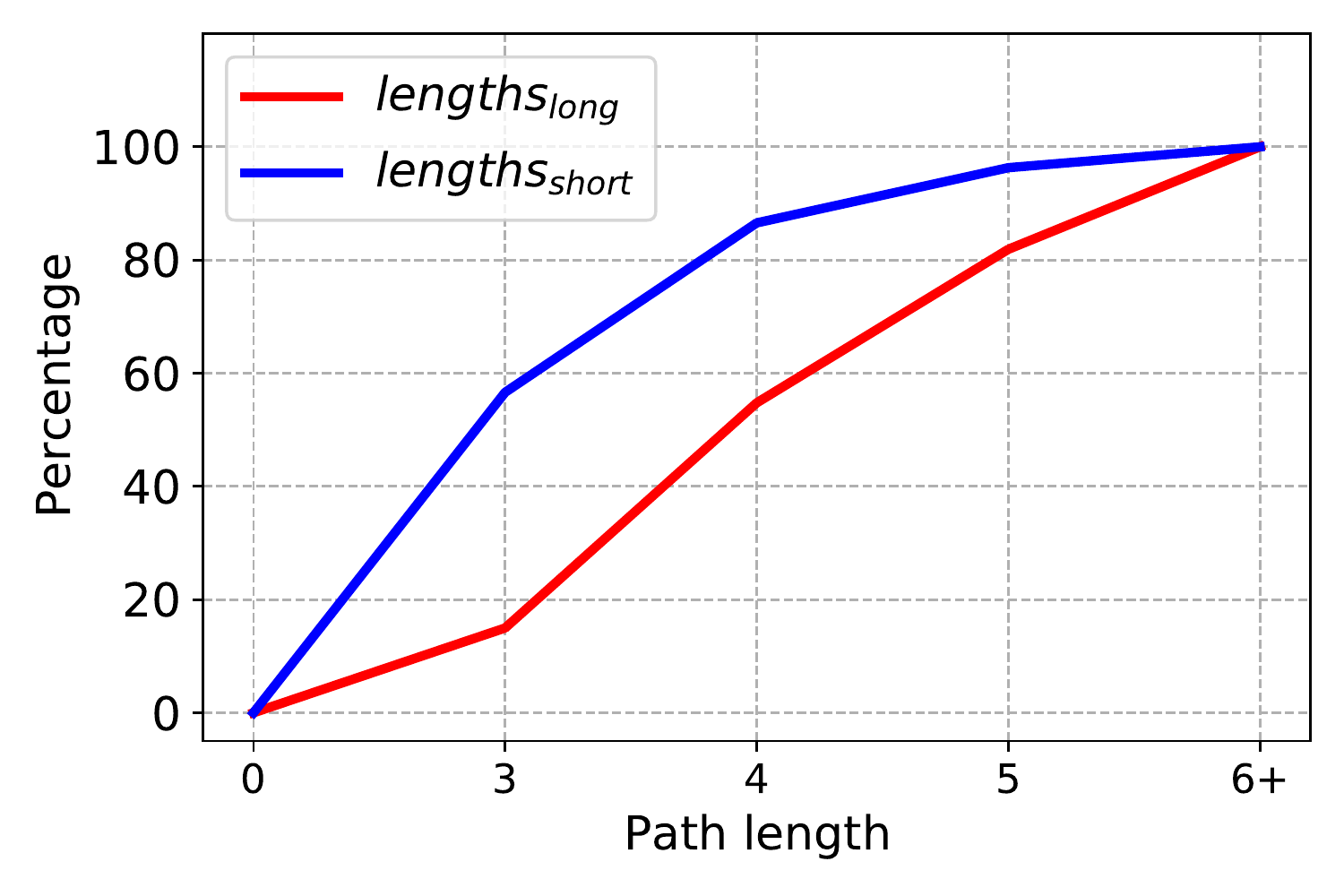}
  \caption{Average path length.}
	\label{fig:paths_length}
\end{figure}

In our first simulation, $\pUB$, our goal was to capture the adversary's
worst case.  This meant we chose $\mathsf{endpoints} =
\textit{uniform}$, so that the choice of sender and receiver was not biased by
connectivity, and thus paths were not short due to their potentially high
connectivity.  Similarly, we chose $\mathsf{values} = \textit{cheap}$ to
minimize the probability of having a payment fail.
For our second simulation, $\pLB$, our goal was to capture the adversary's
best case, so we chose
$\mathsf{endpoints} = \textit{weighted}$ to ensure highly connected nodes
were picked more often and thus paths were shorter.  We also
chose $\mathsf{values} = \textit{expensive}$, leading to many balance failures.

As shown in Figure~\ref{fig:paths_length}, even when we attempted to maximize
the path length in $\pUB$, 14.98\% of paths still consist of only one hop.
In $\pLB$, 56.65\% of paths consisted of a single hop.  This interval agrees
with recent research, which argues that 17-37\% of paths have only one
intermediate node~\cite{beres2019cryptoeconomic}. The main reason the paths
are short even in $\pUB$ is that the network topology and the client path
finding algorithm have a much larger effect on the path length than
$\mathsf{endpoints}$ or $\mathsf{values}$.

\begin{figure}[t!]
  \centering
	\includegraphics[width=0.6\linewidth]{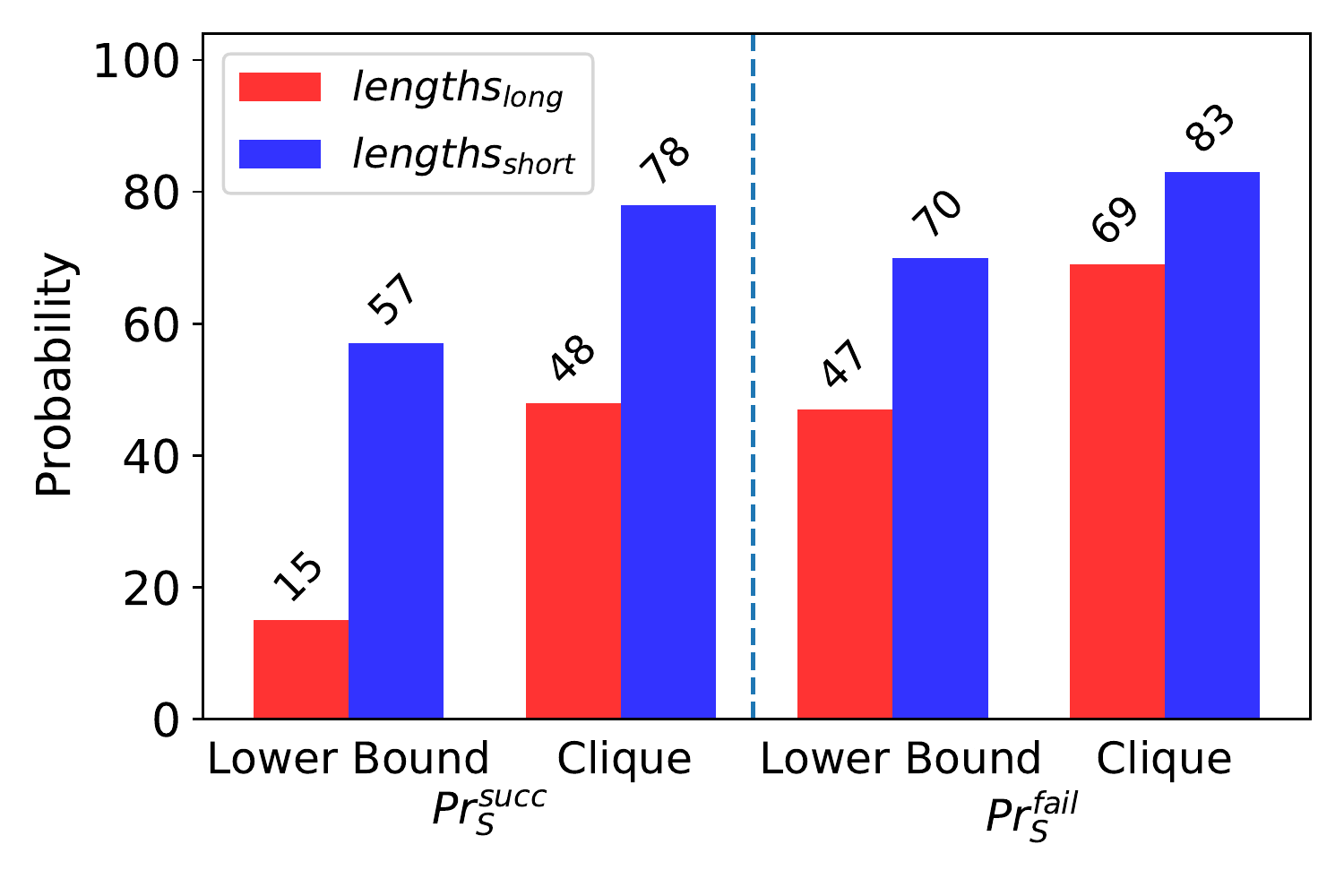}
  \caption{Probability of correctly identifying the sender given successful and
	failed payment. For detailed simulation settings of $\pUB$ and
	$\pLB$ see~\Cref{sec:simulation_results}}
	\label{fig:boat2}
\end{figure}

Beyond the results in Figure~\ref{fig:paths_length}, running our simulator
enabled us to estimate the probabilities
$\Pr{[L = \ell]}$ for %
$3\leq \ell\leq 20$ for both the best- and worst-case scenario for the
adversary.
We now use those results to compute the probabilities $\successprob{S}$ and
$\failprob{S}$ for the case where our adversary is succesful only when it is
impossible to be wrong ($\mathsf{Lower Bound}$) as well as in the case of a
clique topology ($\mathsf{clique}$), as shown in Figure~\ref{fig:boat2}. Here
the clique topology is the worst possible topology for the adversary, since a
less complete topology would allow the adversary to rule out nodes that cannot
be involved in the payment and thus increase their confidence.

$\successprob{S}$ is bounded from below when $L = 3$, since in that case the
adversary can never be wrong. Similarly, $\failprob{S}$ is bounded from below when
$L=4$. In the case of a successful payment, the lower bound on $\successprob{S}$
ranges from $15\%$ ($\pUB$) to $57\%$ ($\pLB$). On the other hand,
the lower bound of $\failprob{S}$ increases with the percentage of unsuccessful
attempts, up to $83\%$ ($\pLB$), which is significantly
higher than any previously recorded experiment. This is also not just a
theoretical result: according to recent measurements, $34\%$ of payments fail
on the first try~\cite{decker}.

Our measurements show that even an adversary following an extremely
simple strategy can have a high probability of inferring the sender of a
payment routed through their node, especially in the case in which the payment
fails.
This is likely due to LN's highly centralized topology, which means paths are
short and often involve the same intermediate nodes, as well as the fact that
clients are designed to find the cheapest---and thus shortest---paths.
Without changes in the client or the network topology, it is thus likely that
intermediate nodes will continue to be able to violate on-path
relationship anonymity.

\section{Payment Discovery}\label{sec:payment_attack}

In this section, we analyze the off-path payment privacy in Lightning,
in terms of the ability of an attacker to learn information about payments
it did not participate in routing.

Informally, our attack works as follows: using the balance discovery attack 
described in Section~\ref{sec:balance_attack}, the attacker constructs
a \emph{network snapshot} at time $t$ consisting of all channels and their
associated balances.  It then runs the attack again at some later time $t+\tau$
and uses the differences between the two snapshots to infer information about
payments that took place by looking at any paths that changed.  In
the simplest case that only a single payment took place between $t$ and $t+\tau$ (and
assuming all fees are zero), the
attacker can see a single path in which the balances changed by some amount
$\amt$ and thus learn everything about this payment: the sender,
the recipient, and the amount $\amt$.
More generally, two payments might overlap in the paths they use, so an
attacker would need to heuristically identify such overlap and
separate the payments accordingly.

\subsection{Payment discovery algorithm}
We define $\tau$ to be the interval in which an attacker is able to capture two
snapshots, $S_t$ and $S_{t+\tau}$, and let $G_{\mathsf{diff}} = S_{t +\tau } - S_{t}$ be
the difference in balance for each channel.
Our goal is then to decompose $G_{\mathsf{diff}}$ into paths representing
distinct payments.
More specifically, we construct paths such that
(1) each edge on the path has the same amount (plus fees),
(2) the union of all paths results in the entire graph $G_{\mathsf{diff}}$, and
(3) the total number of paths is minimal.
This last requirement is to avoid splitting up multi-hop payments: if
there is a payment from $A$ to $C$ along the path $A\rightarrow B\rightarrow
C$, we do not want to count it as two (equal-sized) payments of
the form $A$ to $B$ and $B$ to $C$.

\begin{figure}[t!]
  \centering
    \includegraphics[width=0.6\linewidth]{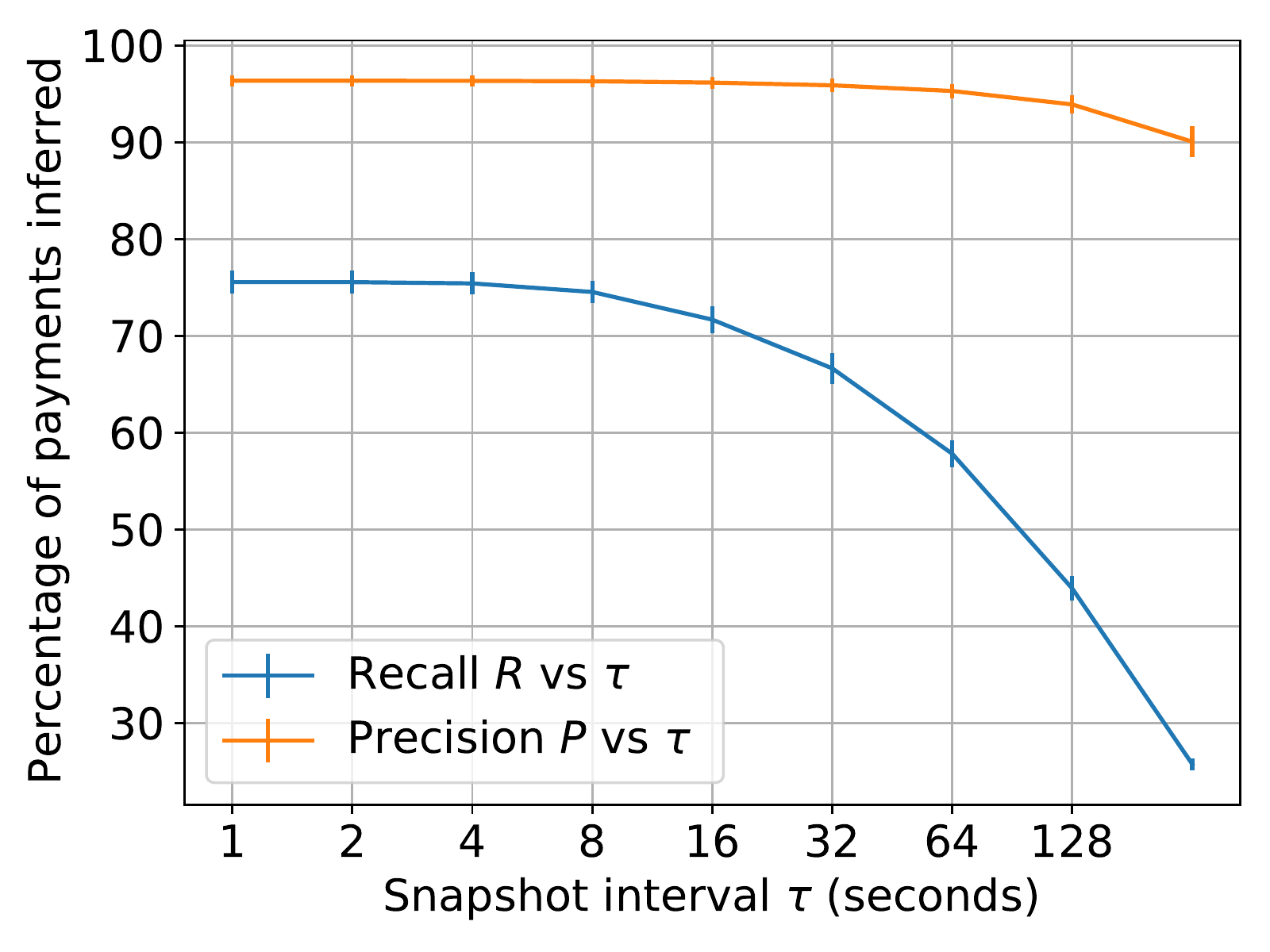}
\caption{The precision and recall of our payment
discovery attack, based on the snapshot interval $\tau$ (in log scale). The error bars show 95\% confidence intervals over five simulation runs.}
    \label{fig:snapshot}
\end{figure}

We give a simple algorithm that solves the above problem under the assumption the
paths are disjoint. This assumption may not always hold, but we will see
in Section~\ref{sec:snapshot_sim_results} that it often holds when the 
interval between snapshots is relatively short.  
Our algorithm proceeds iteratively by ``merging'' payment paths.
We initially consider each non-zero edge in $G_{\mathsf{diff}}$
as a distinct payment.
We then select an arbitrary edge with difference $\amt$,
and merge it with any adjacent edges with the same amount
(plus the publicly known fee $f$)
until no edge of weight $\amt$ can be merged.
\[
A \xrightarrow{\amt+f_{A,B}+f_{B,C}} B,~ B \xrightarrow{\amt+f_{B,C}} C ~\Rightarrow~ \textnormal{infer payment A to C} 
\]
We then remove this path from $G_{\mathsf{diff}}$ and continue with another edge.
Asymptotically the running time of this algorithm is $O(|E|^2)$ for $E$ edges; given
the size and sparsity of Lightning Network today this means it runs in under
a second.

There are several ways this algorithm can make incorrect inferences.
First, it would incorrectly merge two same-valued payments
$A$ to $B$ and $B$ to $C$ occurring end-to-end.
Second, our algorithm does not attempt to resolve the case that a single channel
is used for multiple payments in an interval.
Looking ahead to Section~\ref{sec:snapshot_sim_results}, our experiments show
this happens infrequently when the snapshot intervals are short enough.
Finally, as we saw in Section~\ref{sec:balance_attack}, balance discovery may
fail for some (or many) channels in the network.
Our algorithm takes a conservative approach designed to
minimize the false positive rate: as a final filtering step, it suppresses any
pairs of inferred payments with approximately the same amount (within a small
threshold of two satoshis).

\subsection{Attack simulation}
\label{sec:attack_sim}

We denote the attacker's precision by $\mathsf{P}$ (the number
of correctly detected payments divided by the total number of detected
payments) and recall by $\mathsf{R}$ (the number of correctly
detected payments divided by the number of actual payments).
We are primarily interested in understanding how these performance metrics
depend on the interval at which an attacker takes snapshots ($\tau$), although
we also present in Appendix~\ref{app:snapshot_sim_results_n} an analysis of 
the attacker's recall based on the number of channels it opens ($n$).

To answer these questions we leverage the simulator we developed in Section~\ref{sec:simulator},
and extend it to include the balance discovery attack from
Section~\ref{sec:balance_attack}.
Due to the fact that $98\%$ of the errors in this attack 
were because a node was not online or did not participate in any payments, 
we set a $0.05$ probability of it failing on a functional channel in which both 
nodes are online.  
In keeping with the discussion in Section~\ref{sec:simulating_payments}, we use
$\totalpay = 2000$ as the total number of payments per day and sample the
senders and recipients randomly from all nodes in the network. 
In terms of payment value, our simulation is a pessimistic scenario where the 
payment amounts are very small ($1000$ satoshis on average)
but fluctuate uniformly within a small range around this average ($\pm$10 
satoshis).  This is pessimistic because it is close to the worst-case
scenario, in which all payments have identical amounts, but 
two factors make it likely that real-world payments would have much 
greater variation: (1) payments are denominated in fine-grained units 
($1$ satoshi = $10^{-8}$~BTC), and (2) wallets typically support generating 
payment invoices in units of fiat currency by applying the real-time Bitcoin 
exchange rate, which is volatile.

\subsection{Simulated attack results}
\label{sec:snapshot_sim_results}

In order to figure out the effect of the snapshot interval $\tau$ on 
$\mathsf{P}$ and
$\mathsf{R}$, we take balance inference snapshots of the entire network
for varying time intervals, ranging from $\tau = 1$ second to
$\tau =2^{8}$ seconds.  Each time, we run the simulator for a period of $30$
days, amounting to 60,000 payments in total.
Figure~\ref{fig:snapshot} shows the relationship between $\tau$ and the number
of payments inferred and confirms the intuition that
the attack is less effective the longer the attacker waits between snapshots,
as this causes overlap between multiple payments. At some
point, however, sampling faster and faster offers diminishing returns; e.g.,
for $\tau = 32$ seconds, the attacker
has a recall $\mathsf{R}$ of 66\%, 
which increases slowly to 74.1\% for $\tau = 1$ second. With a realistic minimum 
of $\tau = 30$ seconds, which is the time it took us and others to run the 
balance discovery attack on a single channel~\cite{herrera2019difficulty}),
the attacker has a recall of more than 67\%. Because of our final
filtering step in our discovery algorithm, we have a precision $\mathsf{P}$
very close to 95\% for smaller values of $\tau$.

\section{Conclusions}\label{sec:conclusion}

In this paper, we systematically explored the main privacy properties of
the Lightning Network and showed that, at least in its existing state, each
property is
susceptible to attack.
Unlike previous work that demonstrated similar gaps between
theoretical and achievable privacy in cryptocurrencies, our research does not
rely on patterns of usage or user behavior.  Instead, the same interfaces that
allow users to perform the basic functions of the network, such as connecting
to peers and routing payments, can also be exploited to learn information that
was meant to be kept secret.  This suggests that these limitations may be somewhat
inherent, or at least that avoiding them would require changes at the design
level rather than at the level of individual users.

\section*{Acknowledgements}

George Kappos, Haaroon Yousaf and Sarah Meiklejohn are supported in
part by EPSRC Grant EP/N028104/1, and in part by the EU H2020 TITANIUM
project under grant agreement number 740558. Sanket Kanjalkar and Andrew
Miller are supported by the NSF under agreement numbers 1801369 and
1943499. Sergi Delgado-Segura was partially funded by EPSRC Grant
EP/N028104/1.

\bibliographystyle{plain}
\begingroup
\raggedright
\bibliography{ln_paper.bib}

\begin{thebibliography}{10}

\bibitem{decker}
The lightning conference: Of channels, flows and icebergs talk by christian
  decker.
\newblock \url{https://www.youtube.com/watch?v=zk7hcJDQH-I}.

\bibitem{lightningnetworkbolts}
Lightning network specifications.
\newblock \url{https://github.com/lightningnetwork/lightning-rfc}.

\bibitem{lnbigarticle}
Person behind 40\% of {LN}'s capacity: "{I} have no doubt in {Bitcoin} and the
  {Lightning} {Network}".
\newblock
  \url{https://www.theblockcrypto.com/post/41083/person-behind-40-of-lns-capacity-i-have-no-doubt-in-bitcoin-and-the-lightning-network}.

\bibitem{androulaki2013evaluating}
Elli Androulaki, Ghassan~O Karame, Marc Roeschlin, Tobias Scherer, and Srdjan
  Capkun.
\newblock Evaluating user privacy in {Bitcoin}.
\newblock In {\em International Conference on Financial Cryptography and Data
  Security}. Springer, 2013.

\bibitem{beres2019cryptoeconomic}
Ferenc B{\'e}res, Istvan~Andras Seres, and Andr{\'a}s~A Bencz{\'u}r.
\newblock A cryptoeconomic traffic analysis of {Bitcoins} {Lightning} network.
\newblock {\em arXiv:1911.09432}, 2019.

\bibitem{biryukov2019privacy}
Alex Biryukov, Daniel Feher, and Giuseppe Vitto.
\newblock Privacy aspects and subliminal channels in {Zcash}.
\newblock In {\em Proceedings of the 2019 ACM SIGSAC Conference on Computer and
  Communications Security}, 2019.

\bibitem{biryukov2014bitcoin}
Alex Biryukov, Dmitry Khovratovich, and Ivan Pustogarov.
\newblock Deanonymisation of clients in {Bitcoin} {P2P} network.
\newblock In {\em Proceedings of ACM CCS}, 2014.

\bibitem{bogatyy2019grin}
Ivan Bogatyy.
\newblock Linking 96\% of {Grin} transactions.
\newblock \url{https://github.com/bogatyy/grin-linkability}.

\bibitem{branzei2017charge}
Simina Br{\^a}nzei, Erel Segal-Halevi, and Aviv Zohar.
\newblock How to charge {Lightning}.
\newblock {\em arXiv:1712.10222}, 2017.

\bibitem{conoscenti2018cloth}
Marco Conoscenti, Antonio Vetr{\`o}, Juan~Carlos De~Martin, and Federico Spini.
\newblock The cloth simulator for {HTLC} payment networks with introductory
  lightning network performance results.
\newblock {\em Information}, 2018.

\bibitem{croman2016scaling}
Kyle Croman, Christian Decker, Ittay Eyal, Adem~Efe Gencer, Ari Juels, Ahmed
  Kosba, Andrew Miller, Prateek Saxena, Elaine Shi, Emin~G{\"u}n Sirer, et~al.
\newblock On scaling decentralized blockchains.
\newblock In {\em International Conference on Financial Cryptography and Data
  Security}. Springer, 2016.

\bibitem{danezis2009sphinx}
George Danezis and Ian Goldberg.
\newblock Sphinx: A compact and provably secure mix format.
\newblock In {\em 30th IEEE Symposium on Security and Privacy}, 2009.

\bibitem{engelmann2017towards}
Felix Engelmann, Henning Kopp, Frank Kargl, Florian Glaser, and Christof
  Weinhardt.
\newblock Towards an economic analysis of routing in payment channel networks.
\newblock In {\em Proceedings of the 1st Workshop on Scalable and Resilient
  Infrastructures for Distributed Ledgers}, 2017.

\bibitem{gervais2016security}
Arthur Gervais, Ghassan~O Karame, Karl W{\"u}st, Vasileios Glykantzis, Hubert
  Ritzdorf, and Srdjan Capkun.
\newblock On the security and performance of proof of work blockchains.
\newblock In {\em Proceedings of the 2016 ACM SIGSAC conference on computer and
  communications security}, 2016.

\bibitem{gudgeon2019sok}
Lewis Gudgeon, Pedro Moreno-Sanchez, Stefanie Roos, Patrick McCorry, and Arthur
  Gervais.
\newblock Sok: {Off} the chain transactions.
\newblock {\em IACR Cryptology ePrint Archive}, 2019.

\bibitem{herrera2019difficulty}
Jordi Herrera-Joancomart{\'\i}, Guillermo Navarro-Arribas, Alejandro
  Ranchal-Pedrosa, Cristina P{\'e}rez-Sol{\`a}, and Joaquin Garcia-Alfaro.
\newblock On the difficulty of hiding the balance of {Lightning Network}
  channels.
\newblock In {\em Proceedings of the 2019 ACM Asia Conference on Computer and
  Communications Security (CCS)}, 2019.

\bibitem{HH19}
Abraham Hinteregger and Bernhard Haslhofer.
\newblock Short paper: An empirical analysis of {Monero} cross-chain
  traceability.
\newblock In {\em Proceedings of the 23rd International Conference on Financial
  Cryptography and Data Security (FC)}, 2019.

\bibitem{Kalodner2017BlockSciDA}
Harry~A. Kalodner, Steven Goldfeder, Alishah Chator, Malte M{\"o}ser, and
  Arvind Narayanan.
\newblock Blocksci: {D}esign and applications of a blockchain analysis
  platform.
\newblock {\em arXiv:1709.02489}, 2017.

\bibitem{kappos2018empirical}
George Kappos, Haaroon Yousaf, Mary Maller, and Sarah Meiklejohn.
\newblock An empirical analysis of anonymity in {Zcash}.
\newblock In {\em 27th $USENIX$ Security Symposium '18}.

\bibitem{khalil2017revive}
Rami Khalil and Arthur Gervais.
\newblock Revive: Rebalancing off-blockchain payment networks.
\newblock In {\em Proceedings of the 2017 ACM SIGSAC Conference on Computer and
  Communications Security}, pages 439--453, 2017.

\bibitem{khan2019lightning}
Nida Khan et~al.
\newblock Lightning network: A comparative review of transaction fees and data
  analysis.
\newblock In {\em International Congress on Blockchain and Applications}, pages
  11--18. Springer, 2019.

\bibitem{koshy2014bitcoin}
Philip Koshy, Diana Koshy, and Patrick McDaniel.
\newblock An analysis of anoymity in {Bitcoin} using {P2P} network traffic.
\newblock In {\em International Conference on Financial Cryptography and Data
  Security (FC)}, 2014.

\bibitem{kumar2017traceability}
Amrit Kumar, Cl{\'e}ment Fischer, Shruti Tople, and Prateek Saxena.
\newblock A traceability analysis of {Monero}'s blockchain.
\newblock In {\em European Symposium on Research in Computer Security}.
  Springer, 2017.

\bibitem{malavolta2017concurrency}
Giulio Malavolta, Pedro Moreno-Sanchez, Aniket Kate, Matteo Maffei, and
  Srivatsan Ravi.
\newblock Concurrency and privacy with payment-channel networks.
\newblock In {\em Proceedings of the 2017 ACM SIGSAC Conference on Computer and
  Communications Security}, 2017.

\bibitem{malavolta2019multihop}
Giulio Malavolta, Pedro Moreno-Sanchez, Clara Schneidewind, Aniket Kate, and
  Matteo Maffei.
\newblock Anonymous {Multi-Hop} {Locks} for blockchain scalability and
  interoperability.
\newblock In {\em Proceedings of NDSS}, 2018.

\bibitem{martinazzi2019evolution}
Stefano Martinazzi.
\newblock The evolution of lightning network's topology during its first year
  and the influence over its core values.
\newblock {\em arXiv preprint arXiv:1902.07307}, 2019.

\bibitem{meiklejohn2013fistful}
Sarah Meiklejohn, Marjori Pomarole, Grant Jordan, Kirill Levchenko, Damon
  McCoy, Geoffrey~M Voelker, and Stefan Savage.
\newblock A fistful of {Bitcoins}: characterizing payments among men with no
  names.
\newblock In {\em Proceedings of the 2013 conference on Internet measurement
  conference}. ACM.

\bibitem{moser2018empirical}
Malte M{\"o}ser, Kyle Soska, Ethan Heilman, Kevin Lee, Henry Heffan, Shashvat
  Srivastava, Kyle Hogan, Jason Hennessey, Andrew Miller, Arvind Narayanan,
  et~al.
\newblock An empirical analysis of traceability in the {Monero} blockchain.
\newblock {\em Proceedings on Privacy Enhancing Technologies}, 2018.

\bibitem{nakamoto2019bitcoin}
Satoshi Nakamoto.
\newblock Bitcoin: A peer-to-peer electronic cash system.
\newblock Technical report, Manubot, 2019.

\bibitem{nisslmueller2020active}
Utz Nisslmueller, Klaus-Tycho Foerster, Stefan Schmid, and Christian Decker.
\newblock Toward active and passive confidentiality attacks on cryptocurrency
  off-chain networks, 2020.

\bibitem{nowostawski2019evaluating}
Mariusz Nowostawski and Jardar T{\o}n.
\newblock Evaluating methods for the identification of off-chain transactions
  in the {Lightning Network}.
\newblock {\em Applied Sciences}, 2019.

\bibitem{lockdown}
Cristina P{\'e}rez-Sol{\`a}, Alejandro Ranchal-Pedrosa, Jordi
  Herrera-Joancomart{\'i}, Guillermo Navarro-Arribas, and Joaquin
  Garcia-Alfaro.
\newblock Lockdown: Balance availability attack against lightning network
  channels.
\newblock In Joseph Bonneau and Nadia Heninger, editors, {\em Financial
  Cryptography and Data Security}, pages 245--263, Cham, 2020. Springer
  International Publishing.

\bibitem{poon2016bitcoin}
Joseph Poon and Thaddeus Dryja.
\newblock The {Bitcoin} {Lightning} {Network}: {Scalable} off-chain instant
  payments, 2016.

\bibitem{zcash-anon}
Jeffrey Quesnelle.
\newblock On the linkability of {Zcash} transactions.
\newblock 2017.

\bibitem{reid2013analysis}
Fergal Reid and Martin Harrigan.
\newblock An analysis of anonymity in the {Bitcoin} system.
\newblock In {\em Security and Privacy in Social Networks}. Springer, 2013.

\bibitem{DBLP:journals/tissec/ReiterR98}
Michael~K. Reiter and Aviel~D. Rubin.
\newblock Crowds: Anonymity for web transactions.
\newblock {\em {ACM} Trans. Inf. Syst. Secur.}, 1(1):66--92, 1998.

\bibitem{rohrer2019discharged}
Elias Rohrer, Julian Malliaris, and Florian Tschorsch.
\newblock Discharged payment channels: Quantifying the lightning network's
  resilience to topology-based attacks.
\newblock In {\em 2019 IEEE European Symposium on Security and Privacy
  Workshops (EuroS\&PW)}, pages 347--356. IEEE, 2019.

\bibitem{romiti2020crosslayer}
Matteo Romiti, Friedhelm Victor, Pedro Moreno-Sanchez, Bernhard Haslhofer, and
  Matteo Maffei.
\newblock Cross-layer deanonymization methods in the lightning protocol, 2020.

\bibitem{ron2013quantitative}
Dorit Ron and Adi Shamir.
\newblock Quantitative analysis of the full {Bitcoin} transaction graph.
\newblock In {\em International Conference on Financial Cryptography and Data
  Security}. Springer, 2013.

\bibitem{seres2019topological}
Istv{\'a}n~Andr{\'a}s Seres, L{\'a}szl{\'o} Guly{\'a}s, D{\'a}niel~A Nagy, and
  P{\'e}ter Burcsi.
\newblock Topological analysis of {Bitcoin's} {Lightning Network}.
\newblock {\em arXiv:1901.04972}, 2019.

\bibitem{spagnuolo2014bitiodine}
Michele Spagnuolo, Federico Maggi, and Stefano Zanero.
\newblock Bitiodine: Extracting intelligence from the {Bitcoin} network.
\newblock In {\em International Conference on Financial Cryptography and Data
  Security}. Springer, 2014.

\bibitem{cryptoeprint:2020:303}
Sergei Tikhomirov, Pedro Moreno-Sanchez, and Matteo Maffei.
\newblock A quantitative analysis of security, anonymity and scalability for
  the lightning network.
\newblock Cryptology ePrint Archive, Report 2020/303, 2020.
\newblock \url{https://eprint.iacr.org/2020/303}.

\bibitem{tikhomirov2020probing}
Sergei Tikhomirov, Rene Pickhardt, Alex Biryukov, and Mariusz Nowostawski.
\newblock Probing channel balances in the lightning network, 2020.

\bibitem{tochner2019hijacking}
Saar Tochner, Stefan Schmid, and Aviv Zohar.
\newblock Hijacking routes in payment channel networks: A predictability
  tradeoff.
\newblock {\em arXiv:1909.06890}, 2019.

\bibitem{werman2018avoiding}
Shira Werman and Aviv Zohar.
\newblock Avoiding deadlocks in payment channel networks.
\newblock In {\em Data Privacy Management, Cryptocurrencies and Blockchain
  Technology}, pages 175--187. Springer, 2018.

\bibitem{DBLP:journals/tissec/WrightALS04}
Matthew~K. Wright, Micah Adler, Brian~Neil Levine, and Clay Shields.
\newblock The predecessor attack: An analysis of a threat to anonymous
  communications systems.
\newblock {\em {ACM} Trans. Inf. Syst. Secur.}, 7(4):489--522, 2004.

\bibitem{yen1970algorithm}
Jin~Y Yen.
\newblock An algorithm for finding shortest routes from all source nodes to a
  given destination in general networks.
\newblock {\em Quarterly of Applied Mathematics}, 1970.

\bibitem{yu-monero}
Zuoxia Yu, Man~Ho Au, Jiangshan Yu, Rupen Yang, Quiliang Xu, and Wang~Fat Lau.
\newblock New empirical traceability analysis of {CryptoNote}-style
  blockchains.
\newblock In {\em Proceedings of the 23rd International Conference on Financial
  Cryptography and Data Security (FC)}, 2019.

\bibitem{zhang2019cheapay}
Yuhui Zhang, Dejun Yang, and Guoliang Xue.
\newblock Cheapay: An optimal algorithm for fee minimization in
  blockchain-based payment channel networks.
\newblock In {\em IEEE International Conference on Communications (ICC) 2019}.

\end{thebibliography}
\endgroup

\appendix

\section{More Details on How LN Works}\label{sec:background-app}

\lndetails

\subsection{Hashed time-lock contracts (HTLCs)}\label{sec:htlc}

The previous section describes how the state of a
two-party channel can be updated, in which both Alice and Bob are sure that
they want a payment to go through, and can thus transition to the new state
immediately.  In a broader network of channels, however, it may
very well be the case that two parties need to prepare their channel for an
update that does not happen.  To see why, remember from
Section~\ref{sec:background} that when Alice is picking
a path from herself to Bob, the information she has about the channels along
the way is their capacity $\capacity$, which is $\capacity = \capacity_\inward
+ \capacity_\outward$.  If $\channelid(U_{n-1}\leftrightarrow U_n)$ has
capacity $\capacity \geq \amt$ but $\capacity_\outward < \amt$ (i.e., $U_{n-1}$
does not have enough to pay $U_n$ the amount Alice is asking), then the
payment fails at this point, so no one along the path should have actually
sent any money.  Equally, the payment could fail due to a malicious
intermediary simply deciding not to forward the onion packet or otherwise
follow the protocol.

To thus create an intermediate state, and to unite payments across an entire
path of channels, the Lightning network uses \emph{hashed time-lock contracts},
or \emph{HTLCs} for short.  In using HTLCs, Alice and Bob can still transition
from $\tx_{A,i}$ and $\tx_{B,i}$ to new transactions $\tx_{A,i+1}$ and
$\tx_{B,i+1}$ representing a payment of $n$ coins from Alice to Bob, but the
first message that Alice sends includes the hash $h$ and timeout $t$.  This
signals to Bob to add an additional output $\htlc$ to $\tx_{A,i+1}$, which
sends $n$ coins to Alice if the time is greater than $t$ (as a refund) and
sends them to him if he provides as input a value $x$ such that $H(x) = h$.
Alice and Bob then proceed as usual in exchanging signatures and revocation
keys for their respective transactions.

If at some point before $t$ Bob sends such an $x$ to Alice, then it is clear
to both parties that Bob could claim the $\htlc$ output of $\tx_{A,i+1}$ if it
were posted to the blockchain.  They thus transition to a new state, $i+2$, in
which they go back to two-output commitment transactions, with
the additional $n$ coins now added to Bob's balance.

Going back to the third step in the description provided in
Section~\ref{sec:background} then,
$U_{i-1}$ and $U_i$ prepare their channel by updating to this intermediate
state $j+1$, using the $h$ and $t_i$ provided in the packet $\onion_i$ to form
the $\htlc$ output.  In the fifth and final step, they settle their channel by
updating to state $j+2$ by removing the $\htlc$ output, once $U_i$ has sent
the pre-image $x$ to $U_{i-1}$ and thus demonstrated their ability to claim
it.  If Bob never provides
the pre-image $x$, then by the pre-image resistance of the hash function no
party along the path is able to claim the $\htlc$ output, so the payment
simply does not happen.  If some malicious $U_i$ along the path decides not to
continue forwarding $x$, then all previous $U_k$, $1\leq k\leq i$ can still
claim the $\htlc$ output in the intermediate state.

\section{Node Throughput}\label{sec:throuput_appendix}

\begin{figure}[t!]
  \centering
	\includegraphics[width=0.6\linewidth]{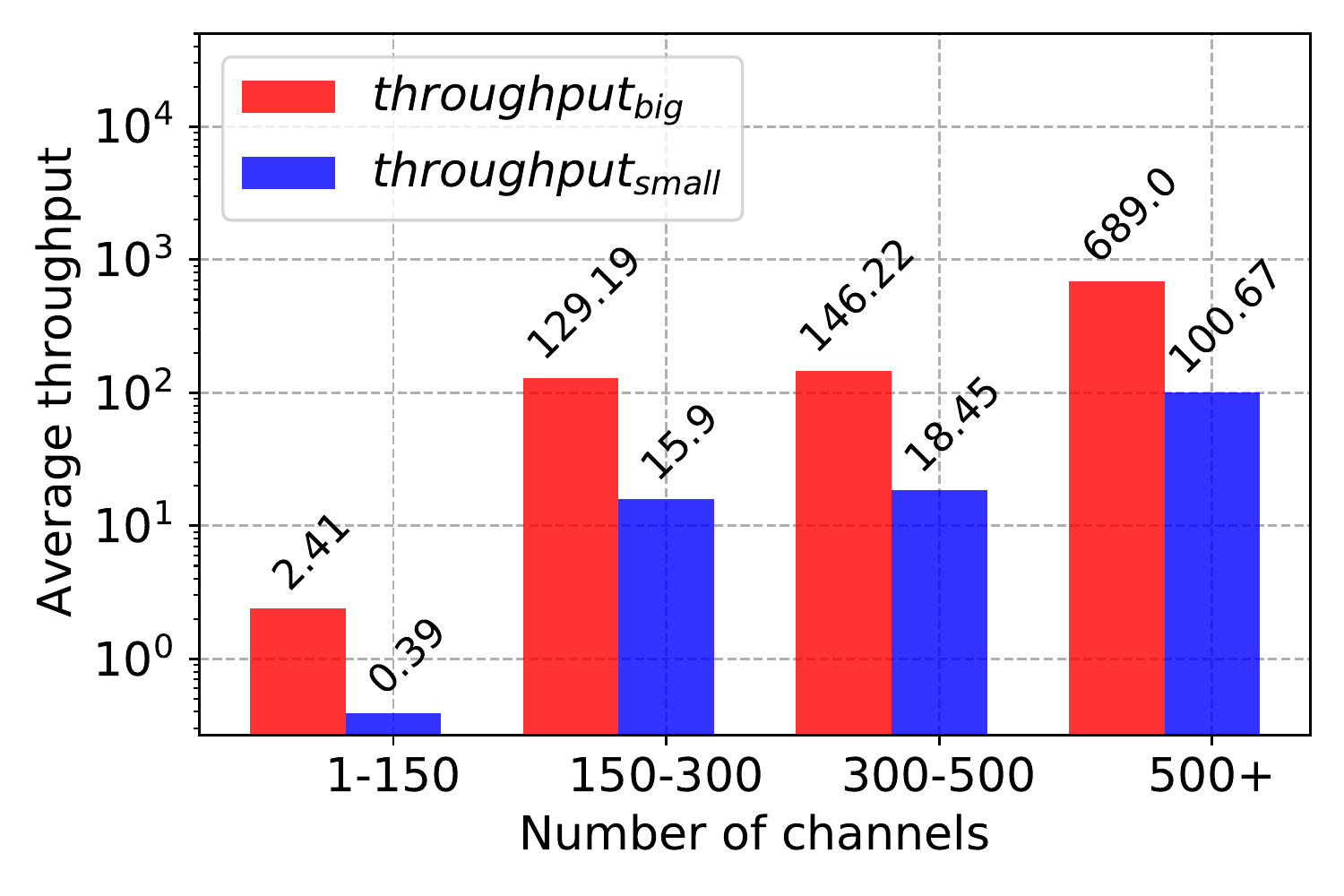}
  \caption{Average throughput}
	\label{fig:throughput}
\end{figure}
Given the parameters $\totalpay$, $\mathsf{endpoints}$, and $\mathsf{values}$, 
we ran two simulation instances,
with the goal of finding the upper and lower bound for the number of payments
a node forwards per day.  Each simulation instance was run using the
network and node parameters scraped on September 1, 2020. We define the throughput
$\throughputpernode$ as the number of payments a node $x$ forwards per day.

In our first simulation instance, $\tUB$, we aim to find the upper bound of
$\throughputpernode$. For this, we use $\totalpay = 10000$ since the throughput
per node increases if the total number of payments increases.
We also use $\mathsf{endpoints} = \textit{weighted}$, which means that
well-connected nodes will be picked more frequently as payment participants.
This will increase $\throughputpernode$ for that set of nodes.
Similarly, we use $\mathsf{values} = \textit{expensive}$, as this again
increases $\throughputpernode$ for nodes that are part of channels with high
capacities.
Using these parameters thus maximizes $\throughputpernode$ for nodes with high
degree and high capacity, which we refer to as \textit{routers}.

In contrast, for $\tLB$, we aim to find the lower bound of
$\throughputpernode$.  We thus use $\totalpay = 1,000$ and $\mathsf{endpoints}
= \textit{uniform}$, as the throughput per node is lowest when the (low)
traffic is spread evenly across them.  Similarly, we use $\mathsf{values} =
\textit{cheap}$ to evenly distribute the load across all channels, even those
with low capacity.

After running the simulations, we group nodes into four
categories, according to the number of channels they possess (1-150, 150-300,
300-500, 500+). We next measure the average number of payments each of these
categories forwards per day.  The results are in Figure~\ref{fig:throughput},
and show that---in both scenarios---the number of channels a node has directly
influences the number of payments it forwards.

In particular, for $\totalpay = 2000$ we see that a
node with over 500 channels forwards around 276 payments per day, whereas a
node with 150-300 channels forwards slightly above 130. When
$\totalpay= 4500$, the node with the maximum number of channels
forwarded 1,485 payments, which translates to a throughput of
0.0172~payments per second.

\vspace{2mm}
\phead{On-path discovery observation}
As mentioned in Section~\ref{sec:introduction}, a node in a path should not
learn anything about the other participating nodes, and even two colluding
nodes should not be able to infer that they belong in the same path. Malavolta
et al.~\cite{malavolta2019multihop} observed that there is a common identifier
in a path (the hash of the secret $\mathbf{x}$) that would make this
trivially possible, but overcame this by blinding the common identifier in
each hop.

In practice, however, as we saw above, even in
$\tUB$, where we maximized the throughput, it still remained low; i.e., 
0.48~payments per minute for the busiest nodes. Given this, if two colluding nodes
are asked to forward a payment of approximately the same value at approximately
the same time (depending on the TCP delays of the nodes in between), it trivial
to infer that they are forwarding the same payment, even with blinded identifiers.
Anonymous communication systems have tackled the same problem of low bandwidth
by adding cover traffic, so LN could similarly add dummy payments to make such
correlation attacks harder to perform. We leave a more thorough exploration of
this countermeasure as future work.

\section{Effect of Attacker Budget on Recall $\mathsf{R}$}
\label{app:snapshot_sim_results_n}

We consider an attacker that chooses the top $n$ nodes with the greatest number of channels.
Here we fix the snapshot interval of
$\tau = 30$ seconds (which, based on our experiment in
Section~\ref{sec:balance_attack},
we believe to be a feasible minimum). In the following discussion, we refer to
our balance discovery attack as a \emph{generic} attack, and to previous
attacks~\cite{herrera2019difficulty,nisslmueller2020active,tikhomirov2020probing}
that rely on error messages as \emph{oracle-aided} attacks.

\begin{figure}[t!]
  \centering
    \includegraphics[width=0.6\linewidth]{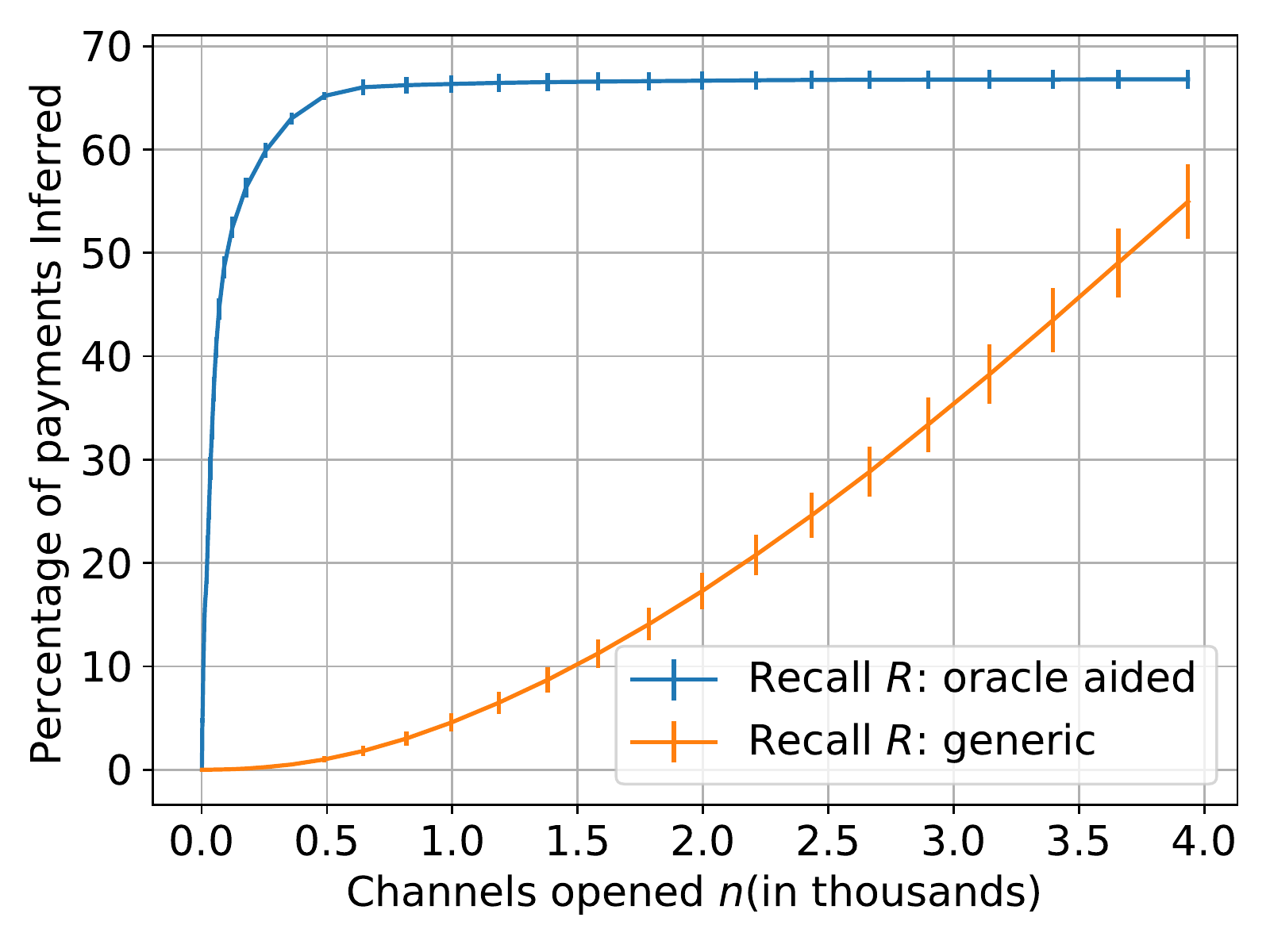}
\caption{Recall $R$ as a function of the number of
channels required. The error bars indicate a 95\% confidence interval
over five simulation runs.}
\label{fig:attacker_cost}
\end{figure}

Figure~\ref{fig:attacker_cost} shows how the number of nodes ($n$) to which the
attacker opens a channel affects the number of payments inferred. For the
same number of nodes attacked,
 we see that the oracle-aided attack performs
significantly
better than the generic attack. This is because the generic attack requires
successfully connecting to both nodes involved in a channel,
whereas either node can suffice for the oracle-aided attack to succeed.
This is also why
we see the recall for the oracle-aided attack
taper off whereas it continues to increase gradually (but with a decreasing
slope) for the generic attack.

As we observe in the oracle-aided attack especially, creating more channels
provides diminishing returns after a certain point. This is expected
given that most payments are routed through
a small subset of nodes, and suggests that if the
attacker is operating with a relatively modest budget, they can choose to attack
a small subset of nodes rather than the entire network. In our simulation, the
oracle-aided attacker achieved a recall of $56.3\%$ after opening only $100$
channels.

Additionally, an attacker may choose to target a subset of nodes, or
even a single target node, for other reasons; e.g., if it is particularly
interested in their payment activities and less interested in the broader
network.
By connecting to and then continuously probing the target node to observe
changes in its balances, the attacker
can identify its payments: if the node was an intermediate routing node, then
the channels around it should have similar incoming and outgoing balances.
If instead only a single channel changed by a particular amount, that node
must have been either the sender or the recipient of a payment of that amount.

\end{document}